\documentclass[annualconference]{acmsiggraph}  

\usepackage{amsfonts}
\usepackage{amsmath}
\usepackage{amssymb}
\usepackage{cases}
\usepackage{wasysym}
\usepackage[ruled,vlined]{algorithm2e}
\usepackage{algorithmicx,algpseudocode}
\usepackage[scaled=.92]{helvet}
\usepackage{times}
\usepackage{paralist}
\usepackage{graphicx}
\usepackage{subfigure}
\usepackage{amssymb}
\usepackage{parskip}
\usepackage[labelfont=bf,textfont=it]{caption}
\usepackage{theorem}
\usepackage{url}
\usepackage{color}
\usepackage{parskip}
\usepackage{listings}
\usepackage{multirow}

\title{Parallel Chen-Han (PCH) Algorithm for Discrete Geodesics}

\author{Xiang Ying~~~~~~Shi-Qing Xin~~~~~~Ying He\thanks{Email: \{ying0008, sqxin, yhe\}@ntu.edu.sg}\\Nanyang Technological University}

\keywords{Discrete geodesic, parallel computation, window
propagation, GPU}

\begin{document}



\maketitle


\begin{abstract}
In many graphics applications, the computation of exact geodesic
distance is very important. However, the high computational cost of
the existing geodesic algorithms means that they are not practical
for large-scale models or time-critical applications. To tackle this
challenge, we propose the parallel Chen-Han (or PCH) algorithm,
which extends the classic Chen-Han (CH) discrete geodesic algorithm
to the parallel setting. The original CH algorithm and its variant
both lack a parallel solution because the windows (a key data
structure that carries the shortest distance in the wavefront
propagation) are maintained in a strict order or a tightly coupled
manner, which means that only one window is processed at a time. We
propose dividing the CH's sequential algorithm into four phases,
window selection, window propagation, data organization, and events
processing so that there is no data dependence or conflicts in each
phase and the operations within each phase can be carried out in
parallel. The proposed PCH algorithm is able to propagate a large
number of windows simultaneously and independently. We also adopt a
simple yet effective strategy to control the total number of
windows. We implement the PCH algorithm on modern GPUs (such as
Nvidia GTX 580) and analyze the performance in detail. The
performance improvement (compared to the sequential algorithms) is
highly consistent with GPU double-precision performance (GFLOPS).
Extensive experiments on real-world models demonstrate an order of
magnitude improvement in execution time compared to the
state-of-the-art.
\end{abstract}

%
%
\keywordlist
%




\section{Introduction}\label{sec:intro}

Computing geodesics on triangle meshes, as a fundamental problem in
geometric modeling, has been widely studied since the mid-1980s.
Over the years, several algorithms have been proposed to compute the
``single-source-all-destination'' geodesic distance, such as, the
MMP algorithm~\cite{Mitchell_Etc:1987}, the CH
algorithm~\cite{Chen_Han:1990} and the improved CH (or ICH)
algorithm~\cite{Xin_Wang:2009}, the fast marching
method~\cite{Kimmel_Sethian:1998} and the approximate MMP
algorithm~\cite{Surazhsky_Etc:2005}. While the state-of-the-art
approaches~\cite{Surazhsky_Etc:2005}~\cite{Xin_Wang:2009} work quite
well for models of moderate size, their high computational cost
means that they are not practical for large-scale models or
time-critical applications.

In the past decade, there has been an increasing trend of performing
the traditionally-CPU-handled computation on a graphics processing
unit (GPU), which uses large numbers of graphics chips to
parallelize the computation. However, developing parallel algorithms
for discrete geodesics is technically challenging due to the lack of
parallel structure; specifically, the geodesic distance is
propagated from the source to all destinations in a sequential
order. To date, the only parallel geodesic algorithm is that of
Weber et al.~\shortcite{DBLP:journals/tog/WeberDBBK08}, who
developed a raster scan-based version of the fast marching
algorithm. Although Weber et al.'s method is highly efficient, it
only computes the first-order approximation of geodesic and requires
parameterization of the surface into a regular domain, which is
usually difficult for surfaces with complicated geometry and/or
topology. To our knowledge, there is no parallel algorithm with
which to compute an exact geodesic on triangle meshes.

\noindent\textbf{Technical challenges}~~The existing exact geodesic
algorithms (for example, the MMP, CH and ICH algorithms) do not
support parallel computing due to the lack of parallel structure.
These algorithms partition each edge into a set of intervals, called
windows, which are maintained in a queue and then propagated across
the mesh faces. The propagation step pops a window from the queue
and then computes its children windows and performs clipping or
merging if necessary, which can add, modify, or remove existing
windows, and the queue is updated accordingly. These algorithms
generate the correct solution regardless of the order in which
windows are removed from the queue; however, selecting windows in
arbitrary order causes extremely slow performance.

The MMP algorithm and the ICH algorithm propagate the windows as a
wavefront by ordering them in a priority queue according to their
distance from the source. However, maintaining a priority queue is a
sequential and global process. It is highly non-trivial to implement
a parallel priority queue on GPUs. The performance of the MMP and
ICH algorithms drops significantly without the priority queue.

The CH algorithm organizes the windows in a tree data structure and
propagates the windows in a breadth-first-search sweep from the root
to the leaves. Due to the data dependency between parents and their
children nodes, only windows on the same level can be processed in
parallel. To maintain this global tree data structure, a CPU/GPU
synchronization must be conducted for each layer. As the depth of
the tree equals the number of mesh faces, the $O(n)$ times
synchronization is too expensive for large models, making parallel
implementation unpractical.
%

\noindent\textbf{Our contributions}~~This paper extends the classic
CH algorithm for parallel computing the exact geodesic distance on
triangle meshes. Our algorithm, which is called the Parallel
Chen-Han (or PCH) algorithm, follows the window propagation
framework, as all the exact algorithms do. Our idea is to divide
CH's sequential algorithm into several phases in such a way that
there is no data dependence in each phase and the operations within
each phase can be done in parallel. Our parallel algorithm design
considers three key factors: (1) propagating the windows in as
parallel a manner as possible; (2) controlling the total number of
windows effectively; and (3) avoiding data conflicts during window
propagation. The following concepts made it possible for us to
address these issues.
\begin{itemize}
\item First, in contrast to the existing approaches, which maintain the
windows in a ``tightly-coupled'' structure (such as the tree data
structure in the CH algorithm) or a ``strictly-ordered'' structure
(such as the priority queue in MMP/ICH algorithms), the windows in
the proposed PCH algorithm are loosely organized. Consequently, our
method neither sorts the windows nor traces the ancestor windows.
Therefore, the windows can be propagated in a fully parallel and
independent manner.

\item Second, to reduce the total number of windows, the PCH algorithm adopts a $k$-selection to determine the $k$
windows nearest to the source, where the parameter $k$ is specified
by the user. These windows will then be processed independently by
multiple GPU threads. Together with a parallel $k$-selection and the
window filtering techniques in the CH and ICH algorithms, our
parallel algorithm is very effective for window propagation, as well
as for controlling the total number of windows. Extensive evaluation
of real-world models shows that our algorithm only has slightly more
windows than the ICH algorithm, which uses a priority queue to
maintain the wavefront.

\item Third, during
window propagation, if a window brings the vertex $v$ a shorter
distance than its current distance, our method does not immediately
update the geodesic distance at $v$ immediately, since other threads
may also access $v$ at the same time and such an update would cause
data conflicts. Instead, our method triggers an update event. Once
all the selected windows have been propagated, our method processes
these events and updates the corresponding data. This updating can
also be done in a parallel manner.
\end{itemize}

The proposed PCH algorithm is the first parallel technique for
calculating exact geodesic distance on triangle meshes. We have
implemented our algorithm on modern GPUs and obtained promising
results on a wide range of models. We have observed that the
performance improvement is highly consistent with the GPUs'
double-precision performance (GFLOPS). For example, our method can
compute the exact geodesic distance field on the 1.8-million-face
Blade model (see Figure 1) in 11 seconds on an Nvidia GTX580
graphics card, which is an order of magnitude faster than the state
of the art~\cite{Xin_Wang:2009}. This indicates that our algorithm
is suitable for applications involving intensive geodesic distance
computations.
\begin{figure} [htbp]
\centering
\includegraphics[width=3.1in]{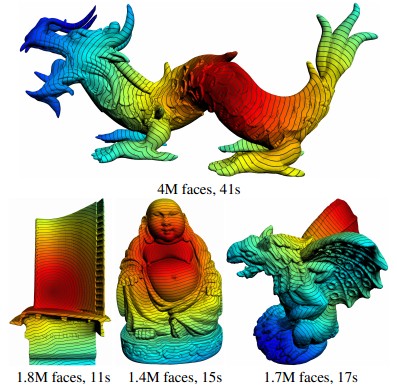}
\label{fig:teaser}\caption{Computing the exact geodesic distance on
triangle meshes by using the proposed PCH algorithm. Our
implementation on Nvidia GTX 580 is one magnitude of order faster
than [Xin and Wang 2009]. }
\end{figure}

\section{Related Work}
\label{sec:related_work} The discrete geodesic problem has been
widely studied since middle 80s. There are many elegant algorithms
to compute the single-source-all-destination geodesic, geodesic
path, geodesic loop and all-pairs geodesic. Due to the limited
space, we review only the representative work of the
single-source-all-destination geodesic, which is our major target.

\noindent\textbf{Exact geodesic algorithms}~~Sharir and
Schorr~\shortcite{Sharir_Schorr:1986} pioneered the discrete
geodesic algorithm with time complexity $O(n^3 \log n)$. However,
their algorithm applies only to convex polyhedral. Using a
continuous Dijkstra sweep, Mitchell, Mount and
Papadimitriou~\shortcite{Mitchell_Etc:1987} presented the first
practical algorithm to compute geodesic distances on general
polyhedral surfaces with time complexity $O(n^2 \log n)$. Liu et
al.~\shortcite{liu} improved the robustness of the MMP algorithm by
presenting effective techniques to handle the degenerate cases.
Surazhsky et al.~\shortcite{Surazhsky_Etc:2005} observed the worst
case running time of the MMP algorithm is rare, and in practice the
algorithm runs in sub-quadratic time. However, the MMP algorithm is
memory inefficient due to its quadratic memory complexity $O(n^2)$,
which diminishes its application to large-scale models. Chen and
Han~\shortcite{Chen_Han:1990} improved the time complexity to
$O(n^2)$ by organizing the windows with a tree data structure. As
only the branch nodes are saved, the CH algorithm has linear space
complexity $O(n)$. Although the time complexity of the CH algorithm
is better than that of the MMP algorithm, extensive experiments show
that the CH algorithm runs much slower than the MMP algorithm,
mainly because most of the windows in the CH algorithm are useless,
which do not contribute to the shortest distance. Xin and
Wang~\shortcite{Xin_Wang:2009} presented an effective window
filtering technique to reduce the windows significantly, and used a
priority queue to order the windows according to the minimal
distance from the source. They demonstrated that their improved CH
algorithm (ICH) outperforms the MMP and CH algorithms in terms of
both execution time and memory cost. Schreiber and
Sharir~\cite{Schreiber:2006} presented an $O(n\log n)$-time
algorithm for convex polyhedral surfaces, which reaches the
theoretical lower bound. However, the optimal time bound for general
polyhedral surfaces is still an open problem. With the exact
geodesic distance, Liu et
al.~\shortcite{Liu:2011:CIB:2006853.2006983} systematically
investigated the analytic structure of iso-contours and bisectors on
triangle meshes, and proposed efficient algorithm for computing
geodesic Voronoi diagrams.

\noindent\textbf{Approximation geodesic algorithms}~~
Sethian~\shortcite{fmm} proposed the $O(n\log n)$ fast marching
method for first-order approximation of geodesics on regular grids.
Kimmel and Sethian~\shortcite{Kimmel_Sethian:1998} extended the fast
marching method to triangle meshes. Weber et
al.~\shortcite{DBLP:journals/tog/WeberDBBK08} presented a raster
scan-based version of the fast marching algorithm for approximating
the geodesic distance on geometry images. Thanks to its parallel
structure, their approach allows highly efficient parallelization on
modern GPUs. Polthier and Schmies~\shortcite{Polthier_Schmies:1998}
proposed the locally straightest geodesic, which differs from the
conventional locally shortest geodesic. Surazhsky et
al.~\shortcite{Surazhsky_Etc:2005} presented the approximate MMP
algorithm that has optimal time complexity $O(n\log n)$ and computes
approximate geodesics with bounded error.


\section{Preliminary}
\label{sec:preliminary} Let $M=(V,E,F)$ be a triangle mesh
representing an orientable 2-manifold where $V$, $E$ and $F$ are the
vertex, edge and face sets, respectively. For a vertex $v\in V$, the
total angle is the sum of interior angles formed between each pair
of neighboring edges incident at $v$. A vertex $v$ is called {\em
spherical} if its total angle is less than $2\pi$, {\em Euclidean}
if the total angle equals $2\pi$, and {\em saddle} if the total
angle is greater than $2\pi$. Mitchell et
al.~\shortcite{Mitchell_Etc:1987} proved that there exists a
geodesic path from the source $s$, typically a mesh vertex, to any
other point on the surface. They also showed that a geodesic path
cannot pass through a spherical vertex except that it is an end
point.

\textbf{Theorem~\cite{Mitchell_Etc:1987}}~~The general form of a
geodesic path is a path that goes through an alternating sequence of
vertices and edges such that the unfolded image of the path along
any edge sequence is a straight line segment and the angle of the
path passing through a vertex is greater than or equal to $\pi$.

If a geodesic path from the source $s$ to destination $t$ passes
through one or more saddle vertices, we call the vertex, which is
nearest to $t$, a \textit{pseudo} source. Clearly, a geodesic path
must be a straight line inside a triangle. When crossing over an
edge, the geodesic path must also correspond to a straight line if
the two adjacent faces are unfolded into a common plane. The exact
geodesic algorithms partition each mesh edge into a set of
intervals, called
\textit{windows}~\cite{Surazhsky_Etc:2005}~\cite{Xin_Wang:2009},
each of which encodes the geodesic paths passing through the
\textit{same} face sequence.

\begin{figure}
\centering
\includegraphics[width=3.1in]{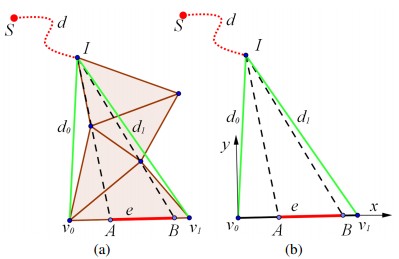}
\caption{(a) A window represents an interval $[A,B]$ on an oriented edge $e$,
which is ``visible'' by the pseudo source $I$ after unfolding the
corresponding face sequence. (b) The position of the pseudo source
on the unfolded plane can be determined by $d_0$ and $d_1$.}
\label{fig:window}
\end{figure}

As shown in Figure~\ref{fig:window}, a window data structure
associated to an edge $e$ is a $6$-tuple $(d, A, B, d_0, d_1, e)$
where
\begin{itemize}
\item $d$ is the distance from the pseudo source to the source
$s$;
\item $A$ and $B$ are the left and right endpoints of the interval;
\item $d_0$ and $d_1$ are the distances from the edge endpoints to the
(pseudo) source.
\end{itemize}

\begin{figure}[htbp]
\centering
\includegraphics[width=3.1in]{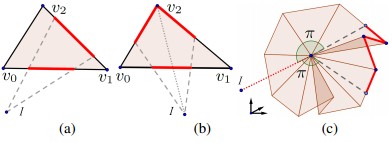}
\caption{Window propagation results in new windows. (a) A window
creates one child on the opposite edge. (b) A window creates two
children on opposite edges. (c) For a saddle point $v$, the
injection ray $l$ is split into two directions (dashed grey lines),
each is an extended straight line of $l$ after unfolding. Then the
new windows are generated on the opposite edges in the fanned area
formed by the two directions. } \label{fig:propagation}
\end{figure}

An important step in the exact geodesic algorithms is to propagate
the windows across the adjacent triangle, which yields new windows
on the opposite edge(s). In general, a window at edge $e$ can have
up to two children at the edges opposite to $e$. See
Figure~\ref{fig:propagation}(a)-(b). For a special case where the
geodesic path passes through a saddle vertex $v$, one or more
windows are then created for the edges in a fanned area. See
Figure~\ref{fig:propagation}(c).

\begin{figure}[htbp]
\centering
\includegraphics[width=3.1in]{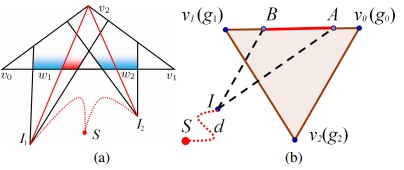}
 \caption{Window filter.
(a) The ``one angle one split'' filter in the CH algorithm. Two
windows $w_1$ and $w_2$ share the same edge $(v_0,v_1)$ and occupy
the vertex $v_2$. However, the pseudo source $I_1$ can not provide a
shorter distance to $v_2$. Thus, only window $w_2$ can generate two
children. (b) The filter used in the ICH algorithm. The window $w$
on edge $(v_0,v_1)$ is generated by some window on edge $(v_1,v_2)$.
Let $g_i$ be the \textit{current} shortest distance for vertex
$v_i$, $i=0,1,2$. The window $w$ is useless, if one of the following
inequalities holds: $d+\|IB\|>g_0+\|v_0B\|$ or
$d+\|IA\|>g_1+\|v_1A\|$ or $d+\|IB\|>g_2+\|v_2B\|$.}
\label{fig:filter}
\end{figure}

Chen and Han~\shortcite{Chen_Han:1990} proposed using a tree data
structure to keep record of the parent-child relationship. The depth
of the tree is equal to the number of faces. To avoid exponential
explosion in the tree, Chen and Han adopted a simple ``one angle one
split'' scheme: if two windows occupy a vertex, at most one of them
can have two children. See Figure~\ref{fig:filter}(a). Xin and
Wang~\shortcite{Xin_Wang:2009} further reduced the number of windows
by using a strict window filtering technique. See
Figure~\ref{fig:filter}(b). They also suggested using the priority
queue to organize the windows according to their distance back to
the source. Adopting the priority queue together with the two window
filtering techniques, the ICH algorithm outperforms both the MMP and
CH algorithms.

Mitchell et al.~\shortcite{Mitchell_Etc:1987} proved that each edge
may have $O(|E|)$ windows and therefore the total number of windows
is $O(|E|^2)$, which is the theoretical upper bound. In practice,
Surazhsky et al.~\shortcite{Surazhsky_Etc:2005} observed that for
typical meshes an edge has an average of $O(\sqrt {|E|})$ windows.
Thus, the key for developing an efficient geodesic algorithm is to
organize the windows effectively.

\section{Computing Geodesics in Parallel}
The proposed parallel Chen-Han algorithm is based on the shared
memory model so that all of the data (such as mesh data structure,
geodesic distance, etc.) are assessable to all processors. This
section presents our algorithm on modern GPUs, as  these are readily
available to the graphics community.

\subsection{Data Structure}

Our algorithm maintains several global variables, as shown in
Table~\ref{tab:variable}. Let $M=(V,E,F)$ denote the input triangle
mesh, where $V$, $E$ and $F$ are the set of vertices, edges and
faces. We use the half-edge data structure to encode $M$'s incidence
information. Considering the limited GPU memory, we adopt a minimal
half-edge data structure. Each edge is decomposed into two
half-edges with opposite orientations. Each half-edge stores the
index of the starting vertex, the index of the opposite half-edge
and its length. Each vertex references one outgoing half-edge.

\begin{verbatim}
struct half_edge {
    int starting_vertex_id;
    int opposite_half_edge_id;
    double length;
};

half_edge he[]; // size 3|F|
half_edge outgoing_he[]; // size |V|
\end{verbatim}

Compared to the standard half-edge data structure, our struct
\textit{half\_edge} does not have the face that it boarders and the
next half-edge around the face. The complete incidence information
is encoded into two arrays, \textit{he} of dimension $3|F|$, which
stores all half-edges of $M$, and \textit{outgoing\_he}, of
dimension $|V|$, which stores one of the half-edges emanating from
each vertex. The three half-edges of the $i$-th triangle are stored
at $he[3i]$, $he[3i+1]$ and $he[3i+2]$, respectively. Given the
\textit{half\_edge}[$j$], the triangle that it boarders is the
$\lfloor j/3\rfloor$-th triangle and the next-half-edge around that
face is \textit{half\_edge}[$3\lfloor j/3\rfloor+(j+1)\%3$], where
$\lfloor \cdot \rfloor$ is the floor function.

To access the one-ring neighbors of the $i$-th vertex $v_i$, one can
simply start with the outgoing half-edge \textit{outgoing\_he}$[i]$
and iteratively switch to the opposite half-edge and find the next
half-edge that points to neighboring vertex. The half-edge data
structure, \textit{he} and \textit{outgoing\_he}, are located in the
GPU memory and accessible to the GPU threads in the real-only
manner. The PCH algorithm does not require the vertex coordinates,
instead, it needs only the metric (the edge length).

\begin{verbatim}
double geod_dist[]; // size |V|
window angle_split[]; // size 3|F|
\end{verbatim}

The array \textit{geod\_dist}[0..$|V|$-1] stores the distance value
at each vertex. The initial value for each non-source vertex is
$\infty$. When the algorithm terminates, the value
\textit{geod\_dist}[\textit{i}] gives the globally shortest geodesic
distance at the $i$-th vertex.

The array \textit{angle\_split}[0..3$|F|$-1] contains the window
that can provide the shortest distance to each angle. Such
information is used by the Chen-Han's ``one angle one split'' window
filter.

During the wavefront propagation, a large number of windows will be
created, processed, and then discarded. To allow efficient access of
the windows, our program creates a memory pool in the GPU memory.
The memory pool contains all the active windows, which are stored
continuously so that there are no memory gaps between adjacent
windows. This condition allows us to distribute the active windows
\textit{evenly} to all GPU threads. The memory pool also provides
each GPU thread with a buffer for the temporary storage of the new
windows. The global array \textit{buffer\_address}[0..$T$-1] stores
the beginning address and size of each buffer. Unlike the continuous
storage of the active windows, these thread buffers are not
physically continuous in the memory pool.

\begin{table}
\begin{tabular}{|l|c|c|c|}
\hline
               Data     & Location            & CPU access    & GPU access\\
\hline \hline

          half\_edge    & GPU RAM             & -    & read only\\
       data structure   &                     &      & \\
  \hline
  \textit{angle\_split}[] & CPU RAM (mapped            & read   & read only\\
  \textit{geod\_dist}[]   & to GPU RAM) & \& write & \\
  \hline memory pool    & GPU RAM             & -             & read \& write\\
  \hline
\end{tabular}
\caption{Global variables.}\label{tab:variable}
\end{table}

\subsection{PCH Algorithm}

Let $S=\{s_i\}_{i=1}^{m}$ be the set of source points specified by
the user. For each source point $s_i$, our algorithm creates a
window for every edge facing $s_i$, and then iteratively processes
the windows.

Each iteration contains the following four steps: First, the
algorithm selects $k$ windows that are nearest to the source points
$s_i$, $i=1,\cdots,m$. Second, the selected $k$ windows are
processed by $T$ threads in a completely independent manner. The
window propagation results in new windows, which are stored in each
thread's own buffer. If a window $w$ can provide a shorter distance
for a vertex $v$, our algorithm does not immediately update the
corresponding entries of $geod\_dist$ or $angle\_split$, as doing so
would cause conflicts. Instead, the algorithm creates an update
event, which will be processed later. Third, the algorithm collects
the new windows and organizes them with the existing windows in the
memory pool, so that there are no memory gaps between the adjacent
windows. Finally, the algorithm processes these update events and
updates the arrays $geod\_dist$ and $angle\_split$ for the
corresponding vertices and corner angles.

\begin{figure}
\centering
\includegraphics[width=3.1in]{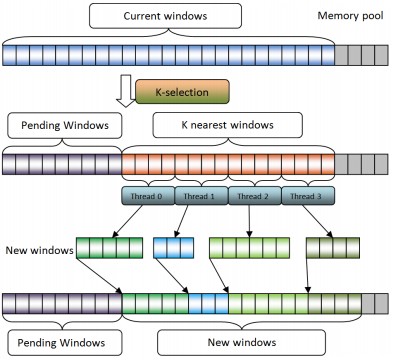}\\
\caption{During each iteration, our algorithm selects $k$ windows
that are nearest to the source points and assigns these windows to
$T$ GPU threads. Each window is taken by one GPU thread and then
propagates independently, which will result in one or more children
windows. After all the selected windows have been processed, our
algorithm collects the newly-generated windows and organize all
windows such that there are no memory gaps between adjacent windows
in the memory pool. The empty slots in the memory pool are drawn in
light gray.} \label{fig:dataflow}
\end{figure}

These four steps are repeated until all windows in the memory pool
have been processed. The data flow for each iteration is shown in
Figure~\ref{fig:dataflow} and the pseudo code is shown in Algorithm
1. Next, we explain the algorithm in details.

\begin{algorithm}
\caption{Parallel Chen-Han Algorithm}
\begin{algorithmic}[1]

\Require A triangle mesh $M=(V,E,F)$, the set of source points
$S=\{s_i|s_i\in V, 1\leq i\leq m\}$, the selection parameter $k$,
and the number of GPU threads $T$

\Ensure The geodesic distance for each vertex

\State initialize the global variables

\State \textcolor{red}{\textbf{parallel}} create a window for every
edge opposite to the source points

\State \textbf{Repeat}

\State ~~~~~\textcolor{red}{\textbf{parallel}} select $k$ nearest
windows

\State ~~~~~\textcolor{red}{\textbf{parallel}} propagate the
selected windows

\State ~~~~~organize the newly-generated windows and update events

\State ~~~~~\textcolor{red}{\textbf{parallel}} process the update
events

\State \textbf{Until} all windows are processed
\end{algorithmic}
\end{algorithm}

\noindent\textbf{Initialization (lines 1, 2)}~~In the initialization
stage, our algorithm creates a memory pool, and two global arrays:
the first array $geod\_dist[|V|]$ contains the geodesic distance at
each vertex, where the initial value is $0$ for the source points
and $\infty$ for all other vertices; the second array
$angle\_split[3|F|]$ contains the windows for each angle, which is
used by the ``one angle one split'' window filter. For every edge
opposite to the source point $s_i$, a window is created and put in
the memory pool.

\noindent\textbf{Step 1. Parallel selection (line 3)}~~Rather than
sorting all windows, our algorithm selects $k$ windows, which are
closest to the source points. The $k$-selection is a fundamental
problem in computer science, which has been widely studied. The
state-of-the art parallel algorithm is due to Monroe et
al.~\shortcite{Monroe2011}. Their algorithm proceeds via an
iterative probabilistic guess-and-check process on pivots for a
three-way partition. When the guess is correct, the problem is
reduced to selection on a much smaller set. They proved that their
probabilistic algorithm always gives a correct result and always
terminates. We adopt Monroe et al.'s parallel algorithm in our GPU
implementation. We also develop an approximate $k$-selection to
improve the speed. We choose $k\gg T$ to fully utilize the pipeline
of the GPU cores. The details of optimal parameter selection and the
approximated version of $k$-selection will be presented in
Sec.~\ref{sec:parameter}.

\noindent\textbf{Step 2. Parallel window propagation (line 4)}~~All
of the selected windows are evenly distributed to the $T$ GPU
threads. Some windows may generate one or more children, while
others may not. Statistics on a wide range of real-world models show
that each window generates an \textit{average} maximum of $3.8$
children\footnote{In general, each window has at most two children
during the propagation. See Figure~\ref{fig:propagation}. However, a
saddle vertex can produce more than 2 windows. The worse case is
that one window is created for each edge opposite to the saddle
vertex. Our statistics show that when processing $L(\geq 8)$ windows
in parallel, the number of new windows is less than $3.8L$.}.
Therefore, we allocate each GPU thread a
$4\lceil\frac{k}{T}\rceil$-sized buffer in the memory pool to
contain the new windows. This fixed-size buffer works well in
practice and we do not observe any buffer overflow in our
experiments. In case of an overflow in the buffer of the $i$-th
thread, for example, the thread $i$ creates a new buffer of doubled
size, copies the contents to the new buffer, and then updates the
entry \textit{buffer\_address}[\textit{i}]. These buffers are not
required to be consecutive in the memory, since they store the new
windows \textit{temporarily}.

\begin{algorithm}[h]
  // assume $w=(d,A,B,d_0,d_1,e)$ is on the oriented\\
  // edge $e=(v_0,v_1)$ and $v_2$ is the opposite vertex of $e$\;
  \If{$A=v_0$ \&\& $d_0+d<geod\_dist[v_0]$}{
    // $w$ provides a short distance to $v_0$\;
    create a distance update event $(v_0,d_0+d)$\;
    \If {$v_0$ is a saddle vertex}{
        create windows in the fanned area (see Fig.~\ref{fig:propagation}(c));
    }
  }
  \If {$B=v_1$ \&\& $d_1+d<geod\_dist[v_1]$} {
    // $w$ provides a short distance to $v_1$\;
    create a distance update event $(v_1,d_1+d)$\;
    \If {$v_1$ is a saddle vertex} {
        create windows in the fanned area (see Fig.~\ref{fig:propagation}(c));
    }
   }
   \If{$w$ occupies $v_2$} {
    // apply the one-angle-one-split filter\;
    \If {$w$ provides $v_2$ a shorter distance than $angle\_split[v_2]$} {
        create an angle update event $(e, w)$\;
        create $w$'s two children (see Fig.~\ref{fig:propagation}(b));
    }
    \Else {
        // $w$ has only one child on edge $(v_0,v_2)$ or $(v_1,v_2)$\;
        create $w$'s child (see Fig.~\ref{fig:filter}(a));
    }
    \If {$\|Iv_2\|+d<geod\_dist[v_2]$}{
        create a distance update event $(v_2, \|Iv_2\|+d)$\;
        \If {$v_2$ is a saddle vertex}{
             create windows in the fanned area (see Fig.~\ref{fig:propagation}(c));
        }
    }
  }
   \Else{
     // $w$ has only one child on edge $(v_0,v_2)$ or $(v_1,v_2)$\;
     create $w$'s child (see Fig.~\ref{fig:propagation}(a));
  }
  apply the ICH filter for the new windows and store the useful\\
  windows in the buffer (see Fig.~\ref{fig:filter}(b));
\caption{PropagatingWindow($w$)}
\end{algorithm}

Note that during the window propagation, if a new window can provide
the vertex $v$ a shorter distance than its current one
$geod\_dist[v]$, we should update $geod\_dist[v]$ accordingly.
However, this would cause data conflicts, as other GPU threads may
also access $v$ at the same time. To avoid conflicts, we create a
\textit{distance} update event for $v$. Similarly, if a window $w$
occupies the vertex $v$ and provides $v$ a shorter distance than
that of the window in the corresponding entry of $angle\_split$, we
also create an \textit{angle} update event. All these events will be
processed later. The pseudo code for window propagation is shown in
function PropagatingWindow().

\noindent\textbf{Step 3. Data organization (line 5)}~~In the above
window propagation step, each GPU thread $T_i$ generates a certain
number of windows, which are stored in $T_i$'s own buffer. The
purpose of this step is to organize these newly-generated windows in
the memory pool such that there are no memory gaps between adjacent
windows, which is a critical condition for the parallel selection
and parallel updating.

Since each GPU thread counts the number of new windows
independently, we use $O(T)$ space to store these numbers, where $T$
is the number of GPU threads. We then compute the accumulated number
of new windows for each thread. Finally, each GPU thread copies its
new windows from its buffer to the memory pool. The pseudo code for
data organization is shown in function OrganizingData(). We use the
same procedure to organize the generated distance and angle update
events.
\begin{algorithm}[h]
\caption{ReorganizeData($\{dataCount\}$, $\{data\}$, $addr$)}
// $data_i$: the buffer storing the output of the $i$-th thread\;
// $dataCount_i$: the size of $data_i$\;
// $addr$: the address of the first free slot in the memory pool\;
$cum_0 \gets 0$ \; \For {$i = 1$ to $T-1$}{
    $cum_i \gets cum_{i-1} + dataCount_{i-1}$
} \textcolor{red}{parallel} \For {each thread $i$}{
    copy $data_i$ to $addr + cum_i$
}
\end{algorithm}

\noindent\textbf{Step 4. Processing the update events (line 6)}~~The window
propagating step produces distance and angle update events, which
are organized in step 3 such that they are stored in the memory pool
in a seamless manner. This allows us to parallel process these
events.

An event is a 2-tuple $(key, value)$. For the distance update event,
$key$ is the vertex id, and $value$ is the geodesic distance. For
the angle update event, $key$ is id of the oriented edge, which
determines the opposite angle/vertex, say $v$, and $value$ is a
window occupies $v$.

To avoid the conflicts in events updating, we must sort all events
by their keys in increasing or decreasing order. If two keys tie, we
compare the two events by their values. For example, for two
distance events $(v_i, d')$ and $(v_i, d'')$ occurring at the same
vertex $v_i$, the one with smaller distance wins. To break a tie
between two windows, we compare the shortest distance from the
source to the corresponding interval. This tie-breaking enables us
to process only the first event at a vertex or angle, since all the
subsequent events occurring at the same vertex or angle do not carry
useful information. We then assign the ordered events evenly to $T$
GPU threads, which update the corresponding vertices or angles. The
pseudo code is shown in function ProcessingEvents().

\begin{algorithm}[h]
\caption{ProcessingEvents($\{event\}$, $\{Data\}$)}
// $event_i$: a 2-tuple $\{key,value\}$\;
// $\{Data\}$: the to-be-updated array $geod\_dist$ or $angle\_split$\;
\textcolor{red}{thrust} sort $\{event\}$ by $key$ and use $value$
for tie breaking\; $S \gets$ size of $\{event\}$\;
\textcolor{red}{parallel} \For {each thread $i$}{
    \For {$j = i \times\lceil\frac{S}{T}\rceil$ to $(i+1) \times\lceil\frac{S}{T}\rceil - 1$} {
        \If {$event_j.key \neq event_{j-1}.key$} {
         $Data[event_j.key] \gets event_j.value$;
        }
    }
}
\end{algorithm}

\subsection{Correctness}
The correctness of our parallel algorithm is based on two
observations. First, during the window propagation, we adopt Chen
and Han's ``one angle one split'' and Xin and Wang's filtering
techniques to reduce the total number of windows. It is
shown~\cite{Chen_Han:1990}\cite{Xin_Wang:2009} that both techniques
remove only the \textit{useless} windows, which do not carry the
shortest distance. Second, Mitchell et
al.~\shortcite{Mitchell_Etc:1987} proved that there exists a
geodesic path from the source $s$ to any other vertex. Our window
propagation step keeps the \textit{useful} windows, which carry the
shortest distance. Thus, for a vertex $v$, we can always find a
sequence of windows $w_1$, $w_2$, $\cdots$, $w_l$, to encode the
geodesic path, where $w_l$ is the last window on the edge $e$
opposite to $v$ such that $w_l$ occupies the vertex $v$. So $w_l$
finally assigns the geodesic distance at $v$. Regardless of the
order in which the windows are processed, such window sequence
always exists. Thus, the resulting distance field is correct.
\section{Implementation \& Experimental Results}

\subsection{Implementation Details} \label{sec:implementation}
We implemented our parallel geodesic algorithm on a 64-bit PC with
an Intel Xeon 2.66GHz CPU and 12GB memory. The graphics card is an
Nvidia GTX 580 with 512 cores and 1.5GB memory. Our program is
compiled using CUDA 4.2.

The input mesh is maintained by using the half-edge data structure.
Only one copy of the mesh structure is transferred to GPU memory.
More specifically, we need only an array of half-edges, each of
which has the length information. We do not need to store the vertex
in the GPU memory; the face connectivity is also encoded in the
half-edge structure.

\begin{figure*}[htbp]
\centering 
\includegraphics[width=6.4in]{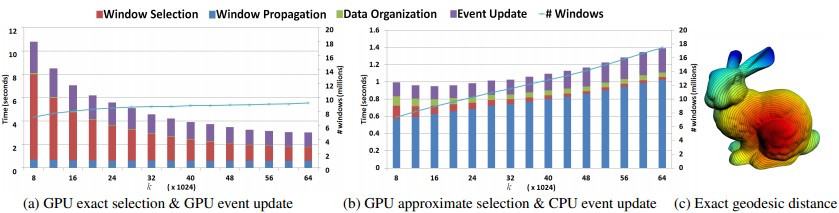}
\caption{The
window selection and event sorting on the GPU take 3 to 6 seconds on
the 144$K$-face Bunny model. We implement an approximate selection
on the GPU and process the events on the CPU, which can
significantly reduce the overall execution time. The vertical axes
are the execution time in seconds and the total number of windows.
The horizontal axes are the selection parameter $k$. Although the
window selection is approximate, the computed geodesic distance is
exact.} \label{fig:approxselection}
\end{figure*}

Our algorithm requires a parallel selection to choose the $k$
nearest windows and a parallel sorting to order all update events by
their keys and values. Selection and sorting can be fully
implemented in parallel. We adopted the probabilistic parallel
selection~\cite{Monroe2011} and CUDA's thrust sorting~\cite{thrust}
in our implementation, but, surprisingly, observed poor performance.
As shown in Figure~\ref{fig:approxselection} (a), the GPU selection
and sorting are very slow for the 144K-face Bunny, taking more than
80\% of the execution time. This slowness is mainly because the GPU
selection algorithms require a large amount of CPU/GPU
synchronization. As a result, these algorithms obtain good
performance only for a very large number of elements (for example,
more than 1 million). To improve the performance, we adopted the
following simple yet effective strategies in our implementation.

First, we implement an approximate $k$-selection on the GPU. Assume
that there are $N (>k)$ windows, $w_1, \cdots, w_N$, in the memory
pool and that there are $T$ threads available. We allocate the
following windows $w_{i}$, $w_{i+T}$, $w_{i+2T}$, $\cdots$, to the
$i$-th thread, which enables each thread to obtain approximately
$\lceil N/T \rceil$ windows, among which the $\lceil k/T\rceil$
nearest windows are selected. Thus, $T$ threads select $k$ windows
in total. This approximate selection uses only a single GPU launch
and can significantly reduce the execution time.

Second, since GPU sorting is not efficient, we copy the update
events from GPU to CPU. To minimize the CPU/GPU data transfer, we
use cudaHostAlloc() to allocate the memory for global variables
$geod\_dist[]$ and $angle\_split[]$. The memory is mapped to CUDA
address space using cudaHostAllocMapped, so both CPU and GPU can
access it. All writing operations on such global variables are
performed by CPU and the GPU procedure PropagatingWindow() accesses
these data in a read-only manner. Although this CPU operation makes
the event updating step sequential, we find it is much more
efficient than the GPU sorting and updating and improves the overall
performance significantly. See Figure~\ref{fig:approxselection}(b).

\noindent\textbf{Remark 1.} Our PCH algorithm requires the windows
to be stored continuously in the physical memory. This condition
plays an important role in the parallel selection step, where the
$i$-th thread checks the $i$, $i+T$, $i+2T$, $\cdots$, positions in
the memory pool, and then selects the window nearest to the source.
The continuously stored windows guarantee that each thread requires
approximately the same number of windows, which means that the
workload is balanced. Also, because different threads may generate
different numbers of windows during the window propagation step, the
procedure ReorganizeData() is required in order to reorganize the
new windows and ensure the condition for the next-round parallel
selection.

\noindent\textbf{Remark 2.} Although our implementation adopts the
\textit{approximation} k-selection to choose the windows, the
correctness of the computed geodesic distance is guaranteed, since
our algorithm does not delete any useful windows and the correctness
of the geodesic distance is independent of the order in which the
windows are processed. Therefore, the resulting geodesic distance is
exact.

\noindent\textbf{Remark 3.} One of the key factors in designing a
parallel algorithm is to effectively avoid data conflicts. Because
our PCH algorithm propagates the windows and then updates the arrays
\textit{geod\_dist} and \textit{angle\_split} in parallel, it is
critical to maintain the data consistency in these arrays. Mutex and
atomic operations are two commonly used techniques to access the
shared memory in parallel computing. However, these techniques are
not suitable to the proposed PCH algorithm, which requires frequent
access of the global arrays \textit{geod\_dist} and
\textit{angle\_split}: each call of PropagatingWindow reads
\textit{geod\_dist} three times and \textit{angle\_dist} once. It
may also write \textit{geod\_dist} up to three times and
\textit{angle\_dist} once.

The large size of both arrays makes it expensive to set a mutex lock
for each element. More importantly, by using mutex lock, each
read/write operation locks the corresponding resources to avoid the
access by other threads, which will significantly compromise the
performance. On the other hand, using the atomic operations in PCH
is not effective either, since doing so when reading/writing the two
global arrays will slow down the procedure PropagatingWindow - the
most expensive of the four steps - thereby compromising the
performance of the entire program.

Therefore, our implementation does not contain the mutex or atomic
operations. Instead, we adopt the delayed update strategy to avoid
data conflicts. Note that the procedure PropagatingWindow does not
update \textit{geod\_dist} and \textit{angle\_dist}. Thanks to the
read-only operations in PropagatingWindow, all the selected $k$
windows can be propagated without any data dependence. Then, in the
ProcessingEvents procedure, we sort the distance and angle updating
events to detect and delete the conflicted data, after which the
arrays can be updated in parallel. This delayed update strategy
works quite well in practice, which is justified by the promising
experimental results.

\begin{figure}[htbp]
\includegraphics[width=3.1in]{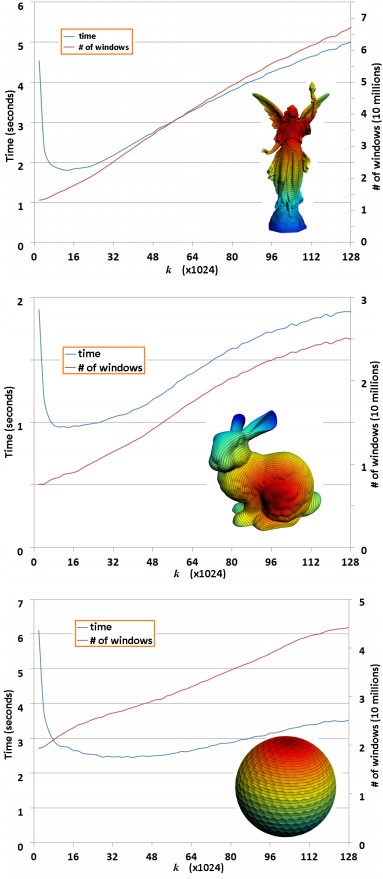}
\caption{Testing the selection parameter $k$ on an Nvidia GTX 580
with 512 cores. The total number of windows (in red) is
approximately linear proportional to the selection parameter $k$
(the horizontal axis). Choosing a small $k$ does not utilize the
GPU's parallel structure at all. On the other hand, choosing a very
large $k$ would produce too many useless windows, resulting in slow
performance (see the blue timing curves). We found the optimal range
for $k$ is $[16\times 2^{10},24\times 2^{10}]$, which leads to the
least execution time. } \label{fig:parameter}
\end{figure}

\subsection{Parameter Setting}\label{sec:parameter}

The performance of our PCH algorithm depends on the mesh complexity
$n=|V|$, the GPU's computational power (such as the number of cores
and frequency), and the selection parameter $k$, which specifies the
number of nearest windows to be processed by GPU threads in
parallel.

The value of $k$ is closely related to the \textit{total} number of
windows produced in the wavefront sweep. Both the MMP algorithm and
the ICH algorithm use a priority queue to determine which window is
nearest to the source. Xin and Wang~\shortcite{Xin_Wang:2009} showed
that the priority queue is very effective in terms of controlling
the total number of windows. If $k=1$, the $k$-selection becomes a
priority queue, leading to a small number of windows. For such a
case, however, our program becomes a sequential program, as only one
window is processed at a time. In order to take full advantage of
the parallel nature of our algorithm, a large $k$ is usually
preferred, so that multiple windows can be processed independently.
On the other hand, for a sufficiently large $k$, the total number of
windows increases dramatically. For example, if $k$ is larger than
the total number of windows, then all windows are selected and
processed simultaneously, which means that the from-near-to-far
wavefront is not maintained at all. As observed in both the MMP and
CH algorithms, such cases will result in an extremely large number
of windows and, therefore, very poor performance. Consequently,
there is a trade-off between controlling the total number of windows
and utilizing the GPU's parallel structure. We tested various $k$ on
a large number of models and observed that the total number of
windows is almost linear to the parameter $k$; the smaller the $k$,
the fewer windows, and vice versa. Figure~\ref{fig:parameter} shows
that the performance curves for Lucy, Bunny and Golf Ball, and the
other models have very similar patterns. We found the optimal range
for $k$ is $[16\times 2^{10},24\times 2^{10}]$, which leads to the
least execution time.

\begin{figure}[htbp]
\centering
\includegraphics[width=3.1in]{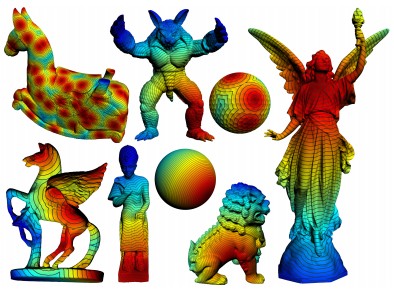}\\
\caption{Given the Bunny model of various resolutions, we found the
optimal range of $k$ is insensitive to the mesh resolution.}
\label{fig:bunny_k_n}
\end{figure}

We also investigated the relationship between the optimal range of
$k$ and the mesh resolution. As Figure~\ref{fig:bunny_k_n} shows,
all the performance curves have the similar pattern, which reveals
the trade-off between utilizing the GPU's parallel structure and
obtaining the optimal performance on the Bunny model of various
resolutions. We found the optimal range of $k$ is insensitive to the
mesh resolution.

\begin{figure}[htbp]
\centering
\includegraphics[width=3.1in]{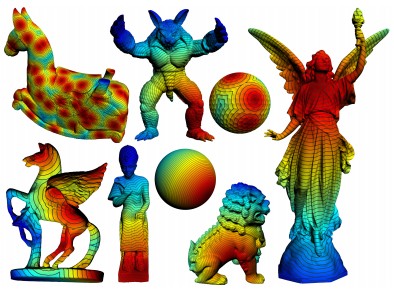}
\caption{More results.} \label{fig:moreresults}
\end{figure}

\begin{figure}[htbp]
\centering
\includegraphics[width=3.1in]{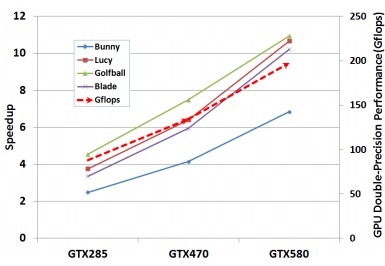}\\
\caption{The performance improvement $T_{ICH}/T_{PCH}$ is consistent
with the GPU double precision performance (GFLOPS).} \label{fig:gpu}
\end{figure}

\subsection{Performance}
We tested our program on a wide range of 3D models, from the small
Bunny (144K faces) to the large-scale Dragon (4 million faces). We
observed that the execution time of the MMP/ICH algorithms and ours
depends on the location of the source point, and that different
source points may result in up to $10\%-15\%$ time difference
because the total number of windows is dependent of the source
point. To obtain a consistent evaluation, we repeated our test 100
times, where a random source point is generated each time.
Table~\ref{tab:speedup} shows the mesh complexity and the
\textit{average} execution time, memory cost, the number of windows,
etc. The consistent experimental results show that our method
obtains an order-of-magnitude improvement for models with more than
500K faces. Our algorithm is also flexible in terms of computing the
``multiple-source-all-destination'' geodesic distance.
Table~\ref{tab:multiple} shows the performance of our algorithm and
the ICH algorithm on the Golf Ball and Pegasus.


\begin{table}[htbp]
\renewcommand{\arraystretch}{1.2}
\centering \setlength{\tabcolsep}{2.0pt}
\begin{small}
\begin{tabular}{|c||c|c|c||c|c|c|}
\hline

\# source & \multicolumn{3}{|c||}{Golf ball} & \multicolumn{3}{|c|}{Pegasus}\\
\cline{2-7}

points    & \begin{small}$T_{ICH}$ \end{small} &
\begin{small} $T_{PCH}$ \end{small} &
\begin{small}$T_{ICH}/T_{PCH}$\end{small} & \begin{small} $T_{ICH}$ \end{small} &
\begin{small} $T_{PCH}$ \end{small} & \begin{small}$T_{ICH}/T_{PCH}$\end{small} \\
\hline

$1 $ & $ 23.71 $ & $ 2.51 $ & $ 9.45 $ & $ 32.61 $ & $ 3.33 $ & $ 9.79$ \\
\hline

$2 $ & $ 22.02 $ & $ 2.30 $ & $ 9.59 $ & $ 33.10 $ & $ 3.04 $ & $ 10.80$ \\
\hline

$4 $ & $ 19.36 $ & $ 2.01 $ & $ 9.65 $ & $ 29.94 $ & $ 2.83 $ & $ 10.56$ \\
\hline

$8 $ & $  14.61 $ & $ 1.46 $ & $ 10.04 $ & $ 34.55 $ & $ 3.16 $ & $ 10.94$ \\
\hline

$16 $ & $ 11.17 $ & $ 1.02 $ & $ 10.90 $ & $ 31.67 $ & $ 3.40 $ & $ 9.33$  \\
\hline

$32 $ & $ 8.49 $ & $ 0.69 $ & $ 12.32 $ & $ 26.02 $ & $ 2.83 $ & $ 9.21 $ \\
\hline

$64 $ & $ 5.99  $ & $ 0.50 $ & $ 11.97 $ & $ 24.33 $ & $ 2.34 $ & $ 10.41$  \\
\hline

$128 $ & $ 4.60 $ & $ 0.37 $ & $ 12.53 $ & $ 20.08 $ & $ 1.99 $ & $ 10.10 $ \\
\hline

$256 $ & $ 3.62 $ & $ 0.29 $ & $ 12.65 $ & $ 15.39 $ & $ 1.57 $ & $ 9.81 $ \\
\hline

$512 $ & $ 2.85 $ & $ 0.23 $ & $ 12.46 $ & $ 12.68 $ & $ 1.25 $ & $ 10.17$ \\
\hline

$1,024 $ & $ 2.36 $ & $ 0.20 $ & $ 12.06 $ & $ 9.95 $ & $ 0.98 $ & $ 10.15$ \\
\hline
\end{tabular}
\end{small}
\caption{Statistics of ``multiple-source-all-destination'' distance.
Execution time was measured in seconds. } \label{tab:multiple}
\end{table}

\begin{table*}
\renewcommand{\arraystretch}{1.2}
\centering \setlength{\tabcolsep}{2.5pt}
\begin{small}
\begin{tabular}{|l|c||c|c||c|c|c||c|c|c|c||c|c|}
\hline
   &   & \multicolumn{2}{|c||}{MMP} & \multicolumn{3}{|c||}{ICH} & \multicolumn{4}{|c||}{PCH} &  \multicolumn{2}{|c|}{Speedup}\\
\cline{3-11}

        &           & Time & Peak &Time  & Peak & Total      & Time & Peak & Total & Peak & \multicolumn{2}{|c|}{$T_{ICH}/T_{PCH}$}\\
        \cline{12-13}

Models  &  \# faces & (s)  & mem. (MB)   & (s)  & mem. (MB)   & \#windows & (s) & mem. (MB) & \# windows & \# windows & Worst & Average \\

 \hline

Bunny     & $144,036$   & $8.19$ & $250.49$ & $6.24$  & $7.04$  &
$6,910,123$ & $0.95$ & $15.10$ & $8,912,758$ & $47,729$ & $5.28$ & $6.57$ \\
\hline

Golf ball & $245,760$   & $26.86$ & $676.32$ & $26.10$ & $12.61$ &
$27,770,244$ & $2.49$ & $22.50$ & $28,541,081$ & $59,218$ & $9.78$ & $10.48$
\\ \hline

Lion & $305,608$ & $17.18$ & $476.36$ & $12.60$ & $14.43$ &
$13,986,984$ & $1.51$ & $26.80$ & $16,315,130$ & $54,916$ & $7.45$ & $8.34$
\\ \hline

Sphere & $327,680$ & $86.94$ & $1,894.02$ & $70.39$ & $21.55$ &
$81,022,375$ & $8.11$ & $37.08$ & $82,412,805$ & $201,499$ & $8.51$ & $8.68$
\\ \hline


Armadillo & $345,944$ & $13.25$ & $412.36$ & $10.34$ & $15.91$ &
$11,169,475$ & $1.40$ & $28.51$ & $13,159,141$ & $35,599$ & $6.04$ & $7.39$
\\ \hline

Lucy & $525,814$ & $22.27$ & $627.03$ & $18.20$ & $22.17$ &
$21,590,268$ & $1.71$ & $42.93$ & $22,996,519$ & $57,691$ & $7.43$ & $10.64$
\\ \hline

Pegasus & $667,474$ & $49.80$ & $1,224.35$ & $33.69$ & $27.47$ &
$28,471,897$ & $3.09$ & $53.15$ & $31,632,899$ & $56,460$ & $8.55$ & $10.90$
\\ \hline

Buddha & $1,439,116$ & $179.67$ & $3,698.71$ & $160.70$ & $67.42$ &
$113,292,750$ & $15.20$ & $114.77$ & $122,402,062$ & $139,761$ & $7.95$ &
$10.57$ \\ \hline

Ramesses & $1,652,528$ & $76.43$ & $1,734.26$ & $52.09$ & $57.90$ &
$47,374,020$ & $5.79$ & $124.72$ & $53,404,855$ & $55,083$ & $7.25$
& $8.99$ \\ \hline

Gargoyle & $1,726,398$ & $211.25$ & $4,096.23$ & $179.90$ & $71.30$ & $119,568,420$ &
$17.50$ & $139.77$ & $137,773,318$ & $208,614$ & $8.47$ & $10.28$ \\ \hline

Blade & $1,765,388$ & $205.91$ & $3,725.16$ & $122.80$ & $72.12$ &
$90,044,841$ & $11.00$ & $137.67$ & $97,250,102$ & $134,014$ & $8.82$ &
$11.16$ \\ \hline

Isidore horse & $2,208,936$ & $80.85$ & $2,026.17$ & $67.77$ &
$69.04$ & $56,403,922$ & $6.31$ & $165.04$ & $68,469,783$ & $54,294$
& $8.95$ & $10.74$
\\ \hline

Dragon & $4,000,000$ & $534.53$ & $8,311.39$ & $421.20$ & $174.37$ &
$265,327,798$ & $41.32$ & $308.40$ & $325,421,513$ & $240,455$ & $8.63$ &
$10.19$ \\ \hline
\end{tabular}
\end{small}
\caption{Model complexity and performance. Tests were repeated 100
times with random source point and the mean values are reported in
this table. The last column shows both the worst and average
speedup. } \label{tab:speedup}
\end{table*}

Figure~\ref{fig:approxselection}(b) shows that GPU processing
 takes more than $75\%$ of the total execution time, while the
 corresponding ratio for large-scale models may exceed $90\%$. As all windows are
processed completely in GPU, the performance of our program depends
heavily on the GPU's parallel computing capacity. We verified this
by testing our program on three Nvidia graphics card: GTX 285, GTX
470 and GTX 580. As Figure~\ref{fig:gpu} shows, we observed that the
performance improvement $T_{ICH}/T_{PCH}$ is highly consistent with
GPU double-precision performance (GFLOPS). Accordingly, we believe
that our algorithm can achieve better performance on the
next-generation graphics card.

In addition to performance, we also measured the peak
memory\footnote{The peak memory of PCH is the actual GPU memory
taken by our algorithm, rather than the size of the memory pool.} of
the MMP, ICH, and PCH algorithms. In order to avoid tiny windows, we
adopt the same tolerance $10^{-6}$ for all the three algorithms. As
Table~\ref{tab:speedup} shows, the memory cost of our algorithm is
much smaller than the MMP algorithm, which has $O(n^2)$ space
complexity. Our memory cost is $1.5-2$ times higher than the ICH
algorithm due to the greater number of windows generated, as well as
the fact that some extra space is required to store the distance and
angle update events.

\section{Comparison \& Discussion}
This section compares our method to the other exact geodesic
algorithms, including the MMP algorithm, the CH algorithm, and the
ICH algorithm. As mentioned before, all these algorithms do not have
parallel structure.

\noindent\textbf{Comparison to the MMP algorithm}~~The MMP algorithm
takes a continuous Dijkstra sweep and maintains the windows on the
wavefront using a priority queue. It is highly non-trivial to
maintain the priority queue in parallel. The MMP algorithm requires
windows clipping during the window propagation. The strong
dependency among the windows makes it difficult to process the
windows in parallel. The MMP algorithm has $O(n^2)$ space
complexity, which diminishes its application on the GPU because the
GPU usually has much less RAM than CPU. Unlike the MMP algorithm,
our method does not require the priority queue and window clipping
operation. Our algorithm also has linear memory complexity.

\noindent\textbf{Comparison to the CH algorithm}~~Instead of using a
priority queue, the CH algorithm organizes the windows in a tree
structure and propagates windows from the root to the leaves.
Because only the branch nodes are retained, the CH algorithm has
linear memory complexity $O(n)$. Chen and Han also proposed a ``one
angle one split'' scheme to avoid exponential explosion. Whenever a
new window $w$ occupies the vertex $v$ and provides $v$ a shorter
distance, the original sub-tree at $v$ is deleted. Due to the
parent-child data dependency, only windows on the same level can be
processed in parallel. As a result, a CPU/GPU synchronization must
be conducted for each layer. Note that the depth of the tree is
equal to the number of mesh faces. Such $O(n)$ synchronization is
very time consuming, which means it is not practical to parallelize
the CH algorithm.

Our algorithm also adopts the ``one angle one split'' scheme to
identify the useless windows, which do not carry the shortest
distance. However, our algorithm neither has the tree structure nor
maintains the parent/child relation. This allows the windows to be
propagated in a fully independent manner.

\noindent\textbf{Comparison to the ICH algorithm}~~The ICH algorithm
is a variant of the CH algorithm, which adopts an effective window
filtering technique to reduce the number of windows. Like the
original CH algorithm, the ICH algorithm maintains the parent-child
relation and the tree structure. Unlike the original algorithm,
however, the window in our algorithm does not contain its parent
information and no windows are sorted. This feature allows us to
propagate a large number of windows in a completely independent
manner.

The ICH algorithm also uses the priority queue to maintain the
wavefront, which is able to minimize the total number of windows.
Instead of the priority queue, our algorithm uses a simple
$k$-selection to determine the windows, which are close to the
source points. This strategy is very effective in terms of
controlling the total number of windows. Our experimental results
show that the total number of windows in our parallel algorithm is
only slightly higher than that of the ICH algorithm.

\noindent\textbf{The PCH algorithm on the CPU}~~We also implement
our PCH algorithm on the CPU in such a way that all the parallel
procedures are replaced by their sequential counterparts. We adopt
quickselection for the sequential $k$-selection, which selects the
top $k$ windows without ordering them, and then compare its
performance to the ICH algorithm on the same hardware setting. We
observe that the CPU PCH algorithm is 5-30\% slower than the ICH
algorithm for all test models.

The priority queue is known to play a critical role in the ICH
algorithm, since it guarantees that the to-be-processed window is
the closest to the source among all active windows. Strictly
ordering the windows by their distances to the source is a very
effective way to control the total number of windows. The CPU PCH
algorithm selects the top $k$ windows and then propagate them one by
one. Because the selected $k$ windows are not ordered, the children
windows of a far-but-early-processed window may be over-ridden by a
close-but-late-processed window. This means that the CPU PCH
algorithm must process process 10-40\% more windows than the ICH
algorithm, resulting in slower performance.

The GPU PCH algorithm processes the same number of windows as the
CPU PCH algorithm. However, the overhead cost of processing the
extra windows is very small compared to the performance speedup by
the high throughput of the GPU. Therefore, the GPU PCH algorithm
significantly outperforms its CPU counterpart and the ICH algorithm.

\noindent\textbf{Window deletion \& complexity analysis} The CH
algorithm organizes the windows in a tree data structure. Chen and
Han~\shortcite{Chen_Han:1990} suggested that once a window becomes
useless,  the useless window and its sub-tree should be deleted (or
tagged) immediately, which can guarantee the total number of windows
is $O(n^2)$, leading to an $O(n^2)$ time complexity. Our PCH
algorithm does not require the tree data structure to maintain the
parent-children relation among windows. Also, it does not delete the
useless windows until the update process. These features
significantly reduce the data dependence, thereby providing great
flexibility in the parallel algorithm design. However, the price to
of such flexibility is the lack of rigorous complexity analysis.

Because time complexity is linearly proportional to the total number
of windows, we empirically evaluated the time complexity by counting
the total number of windows. As Table~\ref{tab:speedup} shows, we
observed that the total number of windows of the PCH algorithm is
only 5\%-20\% more than that of the ICH algorithm. This implies that
such a delay in windows deletion only has a very minor effect on the
time complexity. As Table~\ref{tab:speedup} shows, our PCH algorithm
can achieve 8x-10x speedup for most of the models, which justifies
the good performance of our algorithm.

\section{Conclusion \& Future Work}\label{sec:summary}
It is highly desirable to develop a parallel algorithm to compute
exact geodesic on triangle meshes. However, the existing exact
geodesic algorithms have very strong data dependence and lack the
parallel solution. This paper presents the PCH algorithm, which
extends the classic Chen-Han algorithm to parallel setting. Our idea
is to divide the sequential CH algorithm into four phases, namely
window selection, window propagation, data re-organization and
events processing, such that the operations in each phase have no
dependence or conflicts and can be done in parallel. In contrast to
the original CH algorithm, the proposed PCH algorithm neither
maintains the windows in a tightly coupled manner nor sorts the
windows according to their distance to the source. As a result, it
can process a large number of windows simultaneously and
independently. We also adopt a simple selection strategy which is
effective to control the total number of windows. We implemented our
parallel geodesic algorithm on modern GPUs (e.g., Nvidia GTX 580)
and observed that the performance improvement (compared to the
conventional sequential algorithms) is highly consistent with GPU
double precision performance (GFLOPS), which justifies the parallel
nature of our algorithm. Extensive experiments on real-world models
demonstrate an order of magnitude improvement in execution time
compared to the state-of-the-art.

There are a few interesting future directions. First, we will study
effective windows filtering techniques to further reduce the total
number of windows, thus, improving the overall performance. Second,
our PCH algorithm naturally supports the
multiple-source-all-destination geodesic distance, thus, it has the
potential to improve the performance of the geodesic Voronoi
diagram~\cite{Liu:2011:CIB:2006853.2006983}. Third, as the first
parallel geodesic algorithm with good performance, the PCH algorithm
is highly desired in many time-critical applications which require
extensive computation of geodesics. We will investigate such
applications in the near future.

\bibliographystyle{acmsiggraph}
\bibliography{distancefield_yhe}

\begin{thebibliography}{\protect\citename{Surazhsky et~al\mbox{.} }2005}

\bibitem[\protect\citename{Bell and Hoberock }2011]{thrust}
{\sc Bell, N., and Hoberock, J.}
\newblock 2011.
\newblock {\em GPU Computing Gems}, jade~ed.
\newblock Morgan Kaufmann Publishers, ch.~Thrust: a productivity-oriented
  library for CUDA, 359--371.

\bibitem[\protect\citename{Chen and Han }1990]{Chen_Han:1990}
{\sc Chen, J., and Han, Y.}
\newblock 1990.
\newblock Shortest paths on a polyhedron.
\newblock In {\em Proceedings of Symposium on Computational Geometry},
  360--369.

\bibitem[\protect\citename{Kimmel and Sethian }1998]{Kimmel_Sethian:1998}
{\sc Kimmel, R., and Sethian, J.~A.}
\newblock 1998.
\newblock Computing geodesic paths on manifolds.
\newblock In {\em Proc. Natl. Acad. Sci.}, 8431--8435.

\bibitem[\protect\citename{Liu et~al\mbox{.} }2007]{liu}
{\sc Liu, Y.-J., Zhou, Q.-Y., and Hu, S.-M.}
\newblock 2007.
\newblock Handling degenerate cases in exact geodesic computation on triangle
  meshes.
\newblock {\em The Visual Computer 23}, 9-11, 661--668.

\bibitem[\protect\citename{Liu et~al\mbox{.}
  }2011]{Liu:2011:CIB:2006853.2006983}
{\sc Liu, Y.-J., Chen, Z., and Tang, K.}
\newblock 2011.
\newblock Construction of iso-contours, bisectors, and {V}oronoi diagrams on
  triangulated surfaces.
\newblock {\em IEEE Trans. Pattern Anal. Mach. Intell. 33}, 8, 1502--1517.

\bibitem[\protect\citename{Mitchell et~al\mbox{.} }1987]{Mitchell_Etc:1987}
{\sc Mitchell, J. S.~B., Mount, D.~M., and Papadimitriou, C.~H.}
\newblock 1987.
\newblock The discrete geodesic problem.
\newblock {\em SIAM J. Comput. 16}, 4, 647--668.

\bibitem[\protect\citename{Monroe et~al\mbox{.} }2011]{Monroe2011}
{\sc Monroe, L., Wendelberger, J., and Michalak, S.}
\newblock 2011.
\newblock Randomized selection on the {GPU}.
\newblock In {\em Proceedings of Symposium on High Performance Graphics},
  89--98.

\bibitem[\protect\citename{Polthier and Schmies }1998]{Polthier_Schmies:1998}
{\sc Polthier, K., and Schmies, M.}
\newblock 1998.
\newblock Mathematical Visualization, ch.~Straightest Geodesics on Polyhedral
  Surfaces, 391.

\bibitem[\protect\citename{Schreiber and Sharir }2006]{Schreiber:2006}
{\sc Schreiber, Y., and Sharir, M.}
\newblock 2006.
\newblock An optimal-time algorithm for shortest paths on a convex polytope in
  three dimensions.
\newblock In {\em Proceedings of Symposium on Computational Geometry}, 30--39.

\bibitem[\protect\citename{Sethian }1996]{fmm}
{\sc Sethian, J.}
\newblock 1996.
\newblock A fast marching level set method for monotonically advancing fronts.
\newblock {\em Proc. Nat. Acad. Sci 93}, 4, 1591--1595.

\bibitem[\protect\citename{Sharir and Schorr }1986]{Sharir_Schorr:1986}
{\sc Sharir, M., and Schorr, A.}
\newblock 1986.
\newblock On shortest paths in polyhedral spaces.
\newblock {\em SIAM J. Comput. 15}, 1, 193--215.

\bibitem[\protect\citename{Surazhsky et~al\mbox{.} }2005]{Surazhsky_Etc:2005}
{\sc Surazhsky, V., Surazhsky, T., Kirsanov, D., Gortler, S.~J., and Hoppe, H.}
\newblock 2005.
\newblock Fast exact and approximate geodesics on meshes.
\newblock {\em ACM Trans. Graph. 24}, 3, 553--560.

\bibitem[\protect\citename{Weber et~al\mbox{.}
  }2008]{DBLP:journals/tog/WeberDBBK08}
{\sc Weber, O., Devir, Y.~S., Bronstein, A.~M., Bronstein, M.~M., and Kimmel,
  R.}
\newblock 2008.
\newblock Parallel algorithms for approximation of distance maps on parametric
  surfaces.
\newblock {\em ACM Trans. Graph. 27}, 4.

\bibitem[\protect\citename{Xin and Wang }2009]{Xin_Wang:2009}
{\sc Xin, S.-Q., and Wang, G.-J.}
\newblock 2009.
\newblock Improving {C}hen and {H}an's algorithm on the discrete geodesic
  problem.
\newblock {\em ACM Trans. Graph. 28}, 4, 104:1--104:8.

\end{thebibliography}

\end{document}